\documentclass[12pt,preprint]{aastex}       

\begin{document}
\title{Once-ionized Helium in Superstrong Magnetic Fields}
\author{George G. Pavlov$^1$ and Victor G. Bezchastnov$^{2,3}$}

\affil{$^1$\ Pennsylvania State University,
             525 Davey Lab,
             University Park,
             PA 16802,
             U.S.A.; pavlov@astro.psu.edu}
\affil{$^2$\ Department of Theoretical Astrophysics,
             Ioffe Physical-Technical Institute,
             194021 St.-Petersburg,
             Russia}
\affil{$^3$\ Theoretische Chemie,
             Physikalisch-Chemisches Institut,
             Universit\"at Heidelberg,
             INF 229,
             D-69120 Heidelberg,
             Germany}

\begin{abstract}
It is generally believed that magnetic fields of some neutron stars, 
the so-called magnetars,
are enormously strong, up to $10^{14}$--$10^{15}$ G.
Recent investigations
have shown that the atmospheres of magnetars are 
possibly composed of helium.
We calculate the structure and 
bound-bound radiative transitions
of the He$^+$ ion in superstrong fields, including the effects
caused by the coupling of the ion's internal degrees of freedom        
to its center-of-mass motion.                                
We show that He$^+$ in superstrong magnetic fields can produce
spectral lines with energies of up to $\approx3$ keV, and it may
be responsible for absorption features 
detected
recently in the soft X-ray spectra of several radio-quiet isolated neutron stars.
Quantization of the ion's motion
across                
a magnetic field results in a fine structure of spectral lines, with 
a typical spacing of tens electron-volts in magnetar-scale fields.
It also gives rise to ion cyclotron transitions, whose energies and oscillator
strengths depend on the state of the bound 
ion.
The bound-ion cyclotron lines of He$^+$ can be observed in the 
UV-optical range at $B\lesssim 10^{13}$ G, and they get into the soft 
X-ray range, at $B\gtrsim 10^{14}$ G.
\end{abstract}

\section{Introduction}
It is now widely accepted that some neutron stars (NSs)
have surface magnetic fields as high as 
$10^{14}$--$10^{15}$ G, vs.\ $10^{9}$--$10^{13}$ G in ordinary
radio pulsars.
NSs with such superstrong magnetic fields, dubbed magnetars
(Thompson \& Duncan 1995), include
Anomalous X-ray Pulsars (AXPs) and Soft Gamma Repeaters (SGRs).
In quiescence, the AXPs and SGRs
are radio-quiet X-ray pulsators with periods 
$P=5$--12 s,
characteristic ages $P/2\dot{P}\sim 10^3$--$10^5$ yr,
and 
X-ray luminosities exceeding the spin-down energy loss rate
(in contrast to radio pulsars);
occasionally these objects experience outbursts 
seen in X-rays and/or soft $\gamma$-rays 
(Woods \& Thompson 2006). The hypothesis of superstrong
magnetic fields can explain their bursting behavior and the
quiescent luminosity, and it is consistent with their high 
spin-down rate.
Magnetar-scale fields have also been suggested for the so-called
X-ray-dim Isolated Neutron Stars (XDINSs; e.g., Haberl 2004,
and references therein). Spectra of some of these
sources show puzzling absorption features in a 0.3--0.5 keV range,
which were tentatively interpreted as proton cyclotron lines 
or hydrogen lines in 
magnetic fields of $\sim 10^{14}$ G 
(e.g., van Kerkwijk et al.\ 2004;
Ho \& Lai 2004, and references therein).
Finally, if the magnetic field geometry is strongly different
from that of a centered dipole, the field can reach magnetar-scale
strengths in some regions of
NS surface 
even if the dipole component
is of an ``ordinary'' strength. For instance, Sanwal et al.\ (2002)
detected absorption features
at 0.7 and 1.4 keV in the spectrum of a radio-quiet
isolated NS 1E\,1207.4--5209 and suggested that the lines
could be formed in hot regions of the NS surface with
a magnetic field $\sim 10^{14}$ G. 

Studying spectral lines in NS spectra allows one to measure not only the field strength,
but also the gravitational redshift at the NS surface (hence, the NS mass-to-radius 
ratio), which is of profound importance for constraining the equation
of state and composition of the superdense matter in the NS interiors.
To identify the lines,
one must know the energies
and oscillator strengths of
radiative transitions.
So far, they have been investigated in detail only 
for the simplest hydrogen atom   
(Ruder et al.\ 1994; Potekhin 1994; Pavlov \& Potekhin 1995),  
and first steps have been made in studying multi-electron
atoms (Mori \& Hailey 2002).

We know from X-ray observations
that surface layers of magnetars are rather hot, 
a few million kelvins.
Therefore, their radiation is
likely emitted from gaseous (plasma) atmospheres. 
If there is even a small amount of hydrogen,
$\gtrsim 10^{-20} M_\odot$,
in the NS surface layers
(e.g., deposited by the fallback of an outer envelope
in the course of NS formation),
the properties of the thermal surface
radiation are completely determined by the hydrogen because of 
gravitational sedimentation. 
 However,
the hydrogen can diffuse to deeper layers and burn
into helium and heavier elements
(Chang \& Bildsten 2003,2004).
 According to Chang et al.\ 
(2004),
the rate of diffusive nuclear burning is substantially higher in magnetars,
so that the hydrogen in the magnetar's photosphere is being burned
on a timescale of hours to years, depending on composition of
the underlying material. Consequently, 
atmospheres of 
magnetars 
may be composed of helium rather than hydrogen.

Thus, to 
explore the possibility that NSs with superstrong magnetic fields
have helium atmospheres and interpret observations of these objects,
the properties of helium
and its ions in such fields should be investigated.
The case of fully ionized helium ($\alpha$-particles) is trivial:
it participates in free-free transitions and can produce an ion
cyclotron line at 
$E_{c,\alpha} = 2\hbar eB/(m_\alpha c) = 315\, B_{14}$ eV. 
General properties of the He$^+$ ion in a strong magnetic
field have been studied by Bezchastnov et al.\
(1998;
hereafter BPV98), who calculated the energies
of the so-called tightly-bound states
(see \S2) for $B=1.88\times 10^{13}$ G.
In this Letter, we use the approach developed by BPV98 to study
the energy levels (\S2) and radiative transitions (\S3) in magnetar-scale
fields. Some astrophysical implications of this study are discussed in \S4. 

\section{Energy Levels}
In a magnetic field, the center-of-mass (c.m.) motion of 
an atomic ion 
cannot be separated from the 
internal motion (Schmelcher \& Cederbaum 1991). 
Therefore, to study 
the structure of the moving He$^+$ ion, 
one has to calculate the quantum states of
the {\em two-particle} Hamiltonian 
\begin{equation}
H = -\frac{\hbar^2}{2\mu}\,\frac{\partial^2}{\partial z^2} + H_\perp 
    -\frac{2e^2}{r}~,
\label{H}
\end{equation}
where $z$ and $r$ are the longitudinal (along the magnetic field) and radial 
coordinates of the electron relative to the nucleus, and 
$\mu = m_e m_\alpha/(m_e+m_\alpha)$ is the reduced mass. 
The 
term 
$H_\perp$ describes the motion of 
non-interacting electron and nucleus 
transverse to the field and 
has the energy spectrum 
$E^\perp_{n_-,n_+} = E_{c,e}\left(n_- + 1/2\right) 
+ E_{c,\alpha}\left(n_+ + 1/2\right)$,
where 
$n_-$ and $n_+$ numerate the Landau levels of 
the electron and the nucleus, respectively, and
$E_{c,e} = \hbar eB/(m_e c)$ is the electron cyclotron energy. 
To calculate the quantum states of the Hamiltonian (\ref{H}),
BPV98 employed a coupled-channel approach, in which a channel 
is a common eigenfunction
 of $H_\perp$ and two integrals of motion of the two-particle system:
the square of the transverse component 
of generalized momentum (Avron et al.\ 1981)
and the longitudinal component of the 
angular momentum. 
A bound state of the ion 
can be labeled  
by quantum numbers $s$, $N$, and $\nu$: 
$\nu=0,1,\ldots$ 
determines the longitudinal excitation and $z$-parity,
$\eta=(-1)^\nu$, of the state,
$s=0,1,\ldots$ 
numerates the transverse internal excitations, 
and $N=s,s+1,\ldots$
describes the collective motion of the ion as a whole,
coupled to the 
internal motion.

The energy levels 
in superstrong magnetic fields are shown in
Figures 1 and 2.
We subtract the 
quantity $E^\perp_{0,0}$ from the energies so that the 
boundary 
$E=0$ separates the truly bound states with negative energies from the 
continuum and quasi-bound states with positive energies.
Being plotted as a function of $N$ (Fig.\ 1),
the levels form several ``$s\nu$-branches''.
The branches with $\nu=0$ and $\nu \geq 1$ combine different types of states,
dubbed ``tightly-bound'' and ``hydrogen-like'' states, respectively.

The energies $E_{sN0}$
of the tightly-bound states grow with increasing $N$,
approaching the 
asymptotic values $E=s E_{c,\alpha}$ at $N \to \infty$, so that
the 00-branch approaches the continuum boundary $E=0$,
while the branches with $s\geq 1$ enter the continuum at some finite
$N$, turning into quasi-bound (auto-ionizing) states\footnote{Since
the so-called ``open channels'' were neglected in our calculations,
the results may be rather inaccurate for autoionizing states,
particularly for oscillator strengths.}.
At 
small values of $N$ the energies $E_{sN0}$ grow
with $N$ almost linearly 
and $N$ numerates the cyclotron levels of the bound ion: 
$E_{sN0} = 
E_{ss0} + \hbar \Omega_s (N-s)$, where $\Omega_s = eB/(m_sc)$. 
However, the effective mass of the ion, $m_s$,
is larger than the actual ion's mass,
and it grows with $s$ and $B$  because of the coupling of the 
c.m.\ and 
internal motions.
The growth of $E_{sN0}$ with $N$ slows down
with increasing $N$, and the ion becomes 
decentered 
(the electron cloud 
shifts apart from the nucleus in the transverse plane, 
see BPV98) and weakly bound.
At the intermediate values of $N$ 
the neighboring $s$-branches closely approach one another 
in relatively low magnetic fields; the distance
of the closest approach strongly increases with $B$ so that the
branches become well isolated.

The energies 
of hydrogen-like states ($\nu\geq 1$)
grow very slowly with $N$
which reflects 
increasing decentering of these loosely bound states.
In the magnetic fields of interest, the energies of odd states,
$\nu =1,3,\ldots$, virtually coincide with
the energies of the He$^+$ ion 
in zero magnetic field,
while the energies of even hydrogen-like states lie between
the neighboring odd-$\nu$ energies, approaching the underlying
odd-$\nu$ level with increasing $B$.
Thus, the hydrogen-like levels form narrow
($\Delta E \simeq 54$ eV), almost horizontal strips
below the thresholds $s E_{c,\alpha}$
(only the $s=0$ hydrogen-like branches are shown in Fig.\ 1).
Because these states are less bound and 
more diffuse 
than the tightly-bound states, they can be easier destroyed
by interactions with neighboring particles in a dense medium.

Figure 2 demonstrates the dependence of several level energies on 
magnetic field strength.      
It shows three families of tightly-bound levels $E_{sN0}$ for 
$s=0$, 1, and 2, with $N=s,\ldots, s+4$.
The 0$N$0 states become progressively more bound with increasing $B$,
while the $s\geq 1$ states become quasi-bound at very large $B$.
The number of truly bound states ($E<0$) decreases with increasing $B$,
so that only the 00-branch remains truly bound at $B>7\times 10^{14}$ G.
The thick solid lines show the boundaries of the $s=0$ and $s=1$
continua ($E=s E_{c,\alpha}$), while the lines just below the
continua show the hydrogen-like levels $E_{sN\nu}$ for $s=0$, 1,
and $\nu=1$, 2. 
When the $s\geq 1$ families cross the strips of hydrogen-like levels
(e.g., at $B\simeq 5$--$7\times 10^{14}$ G for $s=1$), the tightly-bound
levels $sN0$ ``repel'' from even hydrogen-like levels $s'N\nu$;
an example of such ``anticrossing'' for $s=1$, $s'=0$, $N=2$, 
$\nu = 2, 4,\ldots 8$
is shown in the inset to Figure 2.
Formally, the number of anticrossings is infinite; we show only the lowest ones,
which results in 
a ``gap'' between the bound states and 
the quasi-bound state 120.

\section{
Radiative Transitions}
A detailed analysis of radiative transitions of He$^+$ in
a magnetic field will be
 presented elsewhere (Bezchastnov \& Pavlov 2005, in preparation).
Here we describe some results of that analysis and provide 
examples of the dependence of transition energies
and oscillator strengths on magnetic field for several
important cases.

In the presence of a magnetic field, the selection rules
and the rates of radiative transitions essentially depend on polarization
of radiation.
In the usual dipole approximation,
the strict selection rules for the transitions $sN\nu \to s'N'\nu'$ are
\begin{equation}
\eta' = (-1)^{\beta+1}\eta, \qquad N'=N+\beta\,,
\end{equation}
where $\eta$ and $\eta'$
are the $z$-parities of the states,  
$\beta=0$ corresponds to the linear polarization along
the magnetic field, and $\beta = +1$ ($-1$) to the right (left)
circular polarization across the magnetic field. 
The account for the c.m.\ motion
invalidates the selection rule $s'=s+\beta$,
applicable in the infinite nucleus mass
approximation,  
which drastically increases the
number of allowed transitions.
In Figure 3,
we show the transition energies
and oscillator strengths
for absorption transitions
from a few
low-lying $0N0$ states.

Transitions
for the longitudinal polarization are only possible between states
with different $z$-parities.
In particular, they can only occur
to odd hydrogen-like levels if the initial state is 
a tightly-bound one. 
Figure 3 shows
the strong $000\to 001$ transition, 
allowed in the infinite nucleus mass
limit, and an example ($010\to 111$) of a weaker transition,
possible only in moving ions.

For the circular polarizations,
there are two different types of transitions.
First, there are transitions between different $s\nu$-branches,
similar to the familiar transitions of the hydrogen atom
in a strong magnetic field (Pavlov \& Potekhin 1995).
For instance, the strongest 
right-polarization transitions from tightly-bound states $0N0$
occur to $s'=1$, $N'=N+1$, $\nu'=0$ (e.g., $000\to 110$
in Fig.\ 3). 
Their transition energies grow with $B$ logarithmically at lower $B$,
and almost linearly at very high $B$. The energy of the main
transition $000\to 110$ reaches about 2.3 keV at
$B=7\times 10^{14}$ G, when the final state becomes quasi-bound.
The oscillator strengths of these transitions are much lower than those
of the linear-polarization transitions, but they may lead to stronger absorption
lines in the spectra of optically thick atmospheres because the lines are
formed in deep layers with large temperature gradients (Zavlin \& Pavlov 2002;
Ho et al.\ 2003). 
Right-polarization transitions with $s'-s >1$
(e.g., $010\to 220$) are allowed for moving
ions,
but they
are much weaker than the $s'=s+1$ transitions.
Left-polarization transitions from the 00-branch, forbidden 
in the infinite nucleus mass limit,
are strongly suppressed
even for moving ions
($020 \to 110$ in Fig.\ 3),
similar to the case of moving hydrogen atoms (Pavlov \& Potekhin 1995).
Also strongly suppressed are circular-polarization transitions
with $\nu'\neq \nu$.

The quantization of collective motion of the He$^+$ ion leads to a
qualitatively new type of bound-bound transitions:
$s'=s$, $\nu'=\nu$, 
$N'-N = +1$ ($-$1)
for absorption (emission) of photons with right (left) circular polarization.
These transitions can be called 
{\em bound-ion cyclotron transitions}
($000\to 010$ in Fig.\ 3). 
At lower $B$, their energies 
($\approx \hbar \Omega_s$)
grow with $B$ linearly,
but the growth slows down
at higher $B$.
The transition energies
are in the observable UV-optical range ($<13.6$ eV) for $B\lesssim 10^{13}$ G,
and in the soft X-ray range ($>0.1$ keV) for $B\gtrsim 10^{14}$ G.

The quantization of He$^+$ motion leads to another interesting 
effect: splitting of a $s\nu\to s'\nu'$ transition into 
a series of $N$-components at energies below that of the ``leading 
component'' corresponding to the transition from $N=0$. 
The splitting is caused by 
different slopes of the $s\nu$-branches
(see Fig.\ 1).
In the low-$N$ regime, the spacing between these
fine structure components is 
$\delta_s \approx \hbar(\Omega_s-\Omega_{s+1})$ 
for right-polarization transitions between tightly-bound levels
(e.g., $\delta_0 \approx 60$ and 130 eV for $B_{14}=1$ and 3,
respectively).
This 
spacing 
can exceed 
collisional and Doppler widths of the fine-structure components.
In this case 
the fine structure 
can be detected in the spectra of NSs with very strong magnetic fields if
the field is sufficiently uniform in the radiating region.
The number of $N$-components,
$\sim kT/\delta_s$, is determined by population 
of the $sN0$ levels. With decreasing $B$,
the spacing decreases, and the number of components
increases,
so that they blend into a broad, asymmetric spectral feature.
Observing the fine structure in longitudinal-polarization transitions is more
problematic because they involve several closely spaced hydrogen-like
levels.

\section{Some Implications}
The results presented above
demonstrate that
transitions between truly bound levels of He$^+$
in magnetar-scale magnetic fields can produce spectral lines
at energies up to $\simeq 3$ keV, and at even higher energies
if auto-ionizing levels are involved.

The broad absorption features at 0.3--0.5 keV observed 
in the spectra of some XDINSs 
(see \S1) 
might be associated with right-polarization transitions in He$^+$
(e.g., $0N0 \to 1,N+1,0$ in $0.4 < B_{14} < 1.5$, at reasonable gravitational
redshifts). The very large widths of the 
features may be
caused by 
overlapping of the
fine structure components. 
To test the suggested interpretation,
deep high-resolution observations should be carried out,
capable to resolve the fine structure of the features
and detect other (weaker) He$^+$,
and perhaps He, spectral lines.

As we mentioned in \S1, if more than one spectral line is detected
and properly identified, one can measure not only the magnetic field
but also the gravitational redshift. For example, 
Sanwal et al.\ (2002) suggested
that the two lines, at 0.7 and 1.4 keV, in the spectrum
of 1E\,1207.4--5259 could be caused by right-polarization and
longitudinal-polarization transitions from the 00-branch of He$^{+}$,
with leading components 000 $\to$ 110 and 000 $\to$ 001,
respectively\footnote{This interpretation is 
incorrect if the two weaker lines, at 2.1 and 2.8 keV, reported
by Bignami et al.\ (2003),
are confirmed
by future observations. Currently, there is some evidence that these
lines are artifacts (Mori et al.\ 2004).
It should also be mentioned that a number of other interpretations
of these two features have been suggested: e.g., He-like oxygen or neon ions
in a magnetic field $\sim 10^{12}$ G (Hailey \& Mori 2002) and hydrogen
molecular ions in $B\sim 4\times 10^{14}$ G (Turbiner \& L\'opez Vieyra
2004). Therefore, the true nature of the lines is currently uncertain. 
}.
Adopting this interpretation, we can first evaluate
the magnetic field for which 
the ratio of the transition energies 
(which does not depend on gravitational
redshift) is $0.7/1.4=0.5$. From the left panel of Figure 3, the ratio of 0.5
corresponds to $B=2.0\times 10^{14}$ G,
and the unredshifted transition energies at this field are
0.9 and 1.8 keV, respectively. The ratio of these
energies to the observed energies is equal to $1+z =
(1-2GM/Rc^2)^{-1/2}= 1.28$,
where $z= 0.28$ is the gravitational redshift, corresponding
to a NS radius 
$R\approx 11\,(M/1.4 M_\odot)$ km\footnote{These values supersede
the crude estimates of Sanwal et al.\ (2002).}.
Interestingly, at this magnetic field and redshift one could expect
an absorption feature at about 2 keV caused by right-polarization
photoionization transitions from the ground state to the 
$s=1$ continuum 
(above the upper bold line in Fig.\ 2).
A feature at $\approx 2$ keV was noticed in the {\sl Chandra}
and {\sl XMM-Newton} observations (Sanwal et al.\ 2002;
Mereghetti et al.\ 2002), but, unfortunately, there are 
poorly calibrated features in the telescope responses at this energy.
This interpretation could be confirmed
if observations with
high energy resolution resolve the predicted
fine structure of the 0.7 keV feature (such an observation
is included in the {\sl Astro-E2} program)
and detect other He$^+$ features. Another
prediction of this interpretation, that the radiation 
is orthogonally polarized in the two features, could be
tested in polarimetric X-ray observations.

To conclude, our results demonstrate that the spectral features
observed in the soft X-ray spectra of some NSs can be produced
by bound-bound transitions of He$^+$ in superstrong magnetic fields.
The next step in understanding the magnetar spectra would be
the study of continuum and, in particular, quasi-bound states 
of He$^+$ and the corresponding radiative transitions. 
\acknowledgements
We thank the anonymous referee whose remarks helped us to improve the presentation.
This work was partially supported by SAO/NASA grants
GO3-4091X and GO5-6074A, and
NASA LTSA grant NAG5-10865.


\vfill\eject
\begin{figure}[!th]
\vspace{-2cm}
\includegraphics[width=19cm]{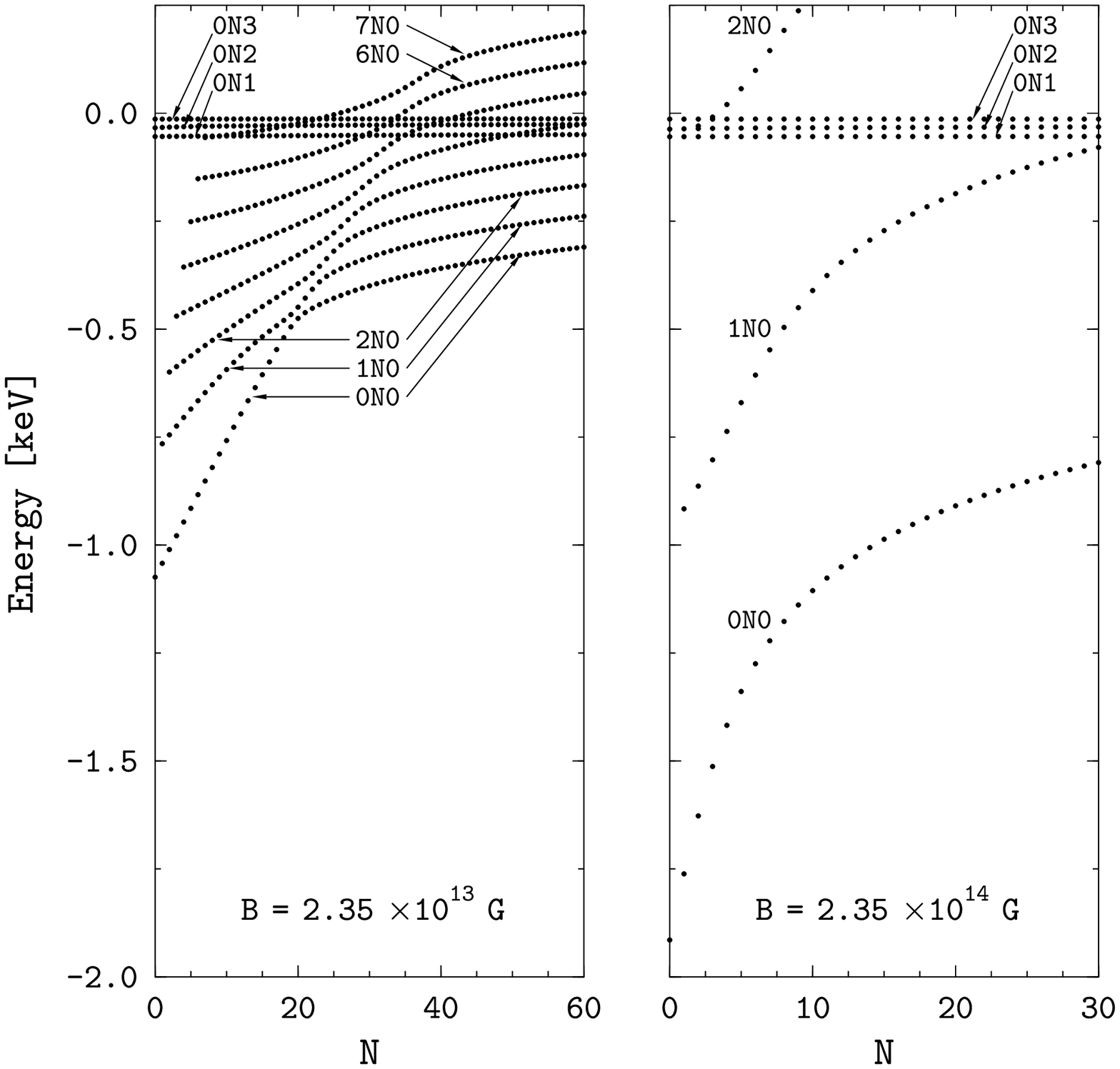}
\caption{
Energy levels $E_{sN\nu}$
vs.\ quantum number $N$ that describes
the quantized motion of the ion as a whole,
for two values of magnetic field.
The labels $sN\nu$ denote $s\nu$-branches.
}
\end{figure}
\begin{figure}
\vspace{-2cm}
\includegraphics[width=16cm]{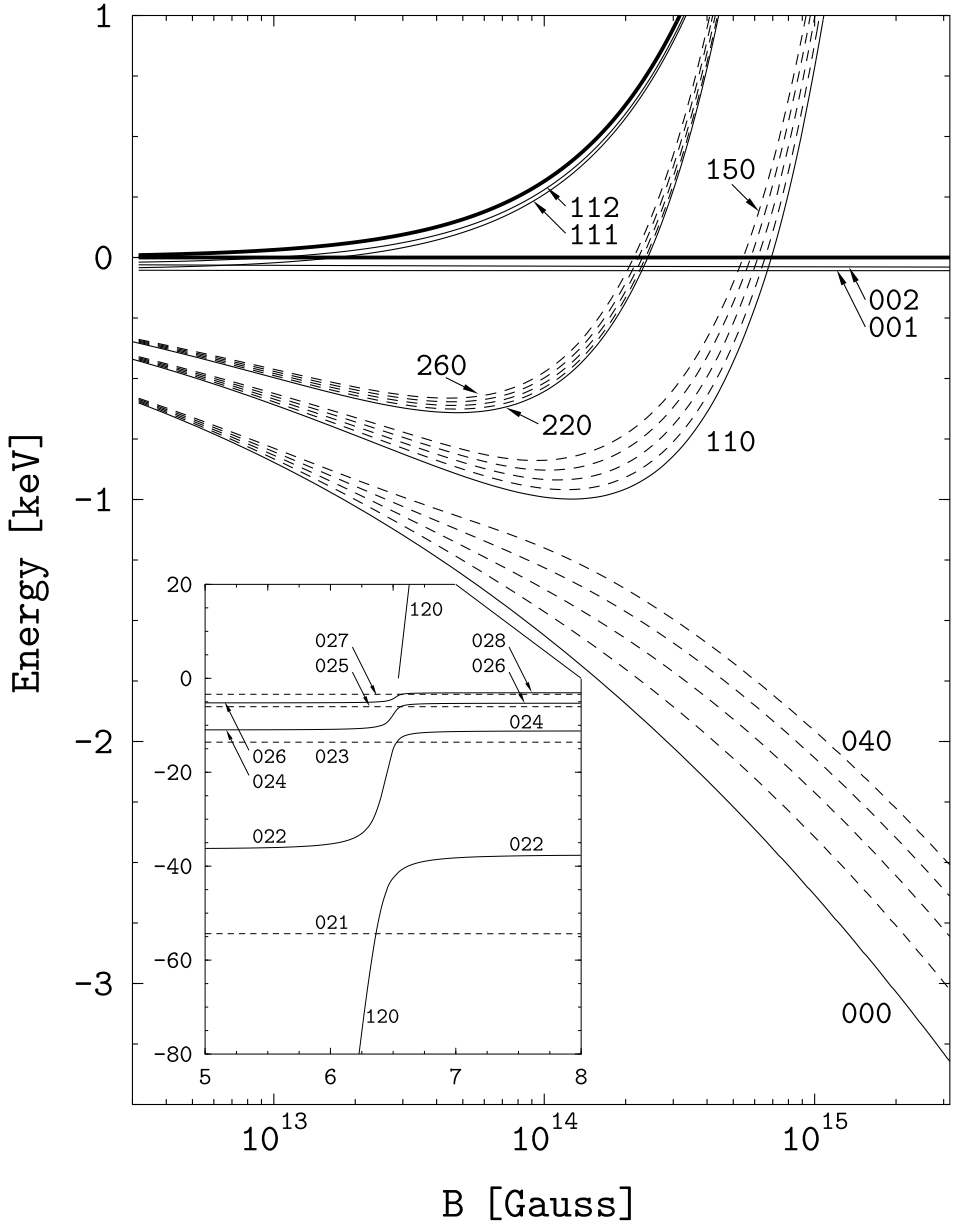}
\caption{
Dependence of the energy
levels
on magnetic field strength.                             
Solid lines show the lowest levels with $N=s$, while the    
dashed lines show the motionally excited levels with $N>s$. 
Two bold solid lines show the 
$s=0$ and $s=1$ continua
thresholds,
$E=0$ and $E=E_{c,\alpha}$, respectively.
The inset demonstrates the anticrossings of the 120 level with lowest
even hydrogen-like levels 02$\nu$ around $B = 6.5\times 10^{14}$ G.
The units of magnetic field and energy for the inset
are $10^{14}$ G and eV,
respectively.                                            
}
\end{figure}
\begin{figure}
\vspace{-5cm}
\includegraphics[width=18cm]{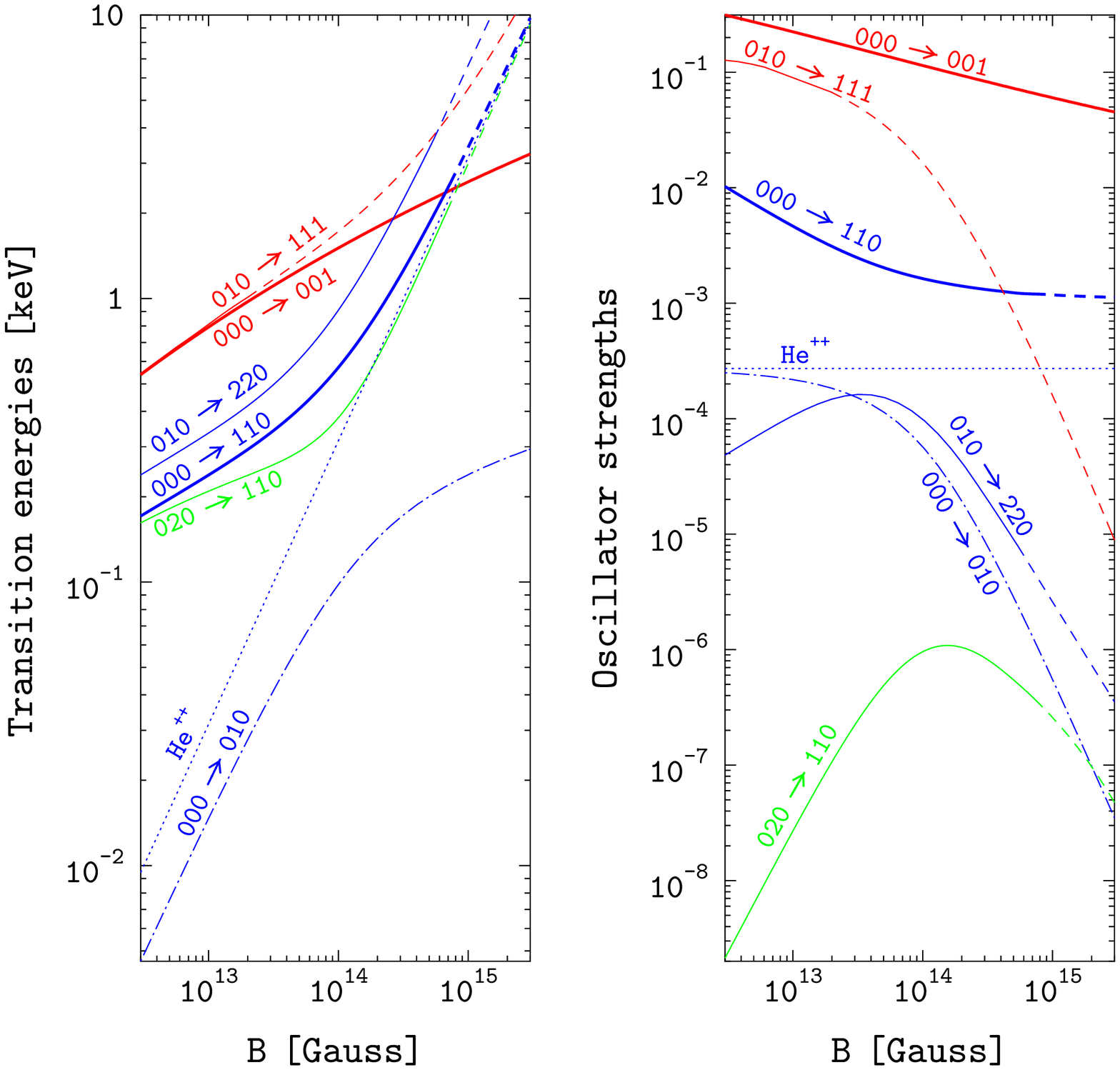}
\caption{
Transition energies and radiative strengths for
several
$0N0 \to s'N'\nu'$ transitions,
for linear polarization along the magnetic field (red curves),
and right (blue)
and left (green)
circular polarizations perpendicular to the
magnetic field.
Transitions allowed in the infinite nucleus mass
limit are shown by bold lines. Dashed parts of the curves
correspond to transitions to auto-ionizing states.
An example of the bound-ion cyclotron transition
is shown by dash-dotted curves.
The dotted lines correspond to the cyclotron
transitions of He$^{++}$.
}
\end{figure}


\begin{thebibliography}{}

\bibitem{1} 
Avron, J.\ E., Herbst, I.\ W., \& Simon, B. 1981, Commun.\ Math.\ Phys., 79, 529

\bibitem{2}
Bezchastnov, V.\ G., Pavlov, G.\ G., \& Ventura, J. 1998, Phys.\ Rev. A,
58, 180 (BPV98)

\bibitem{3}
Bignami, G.\ F., Caraveo, P.\ A., De Luca, A., \& Mereghetti, S. 2003, Nature, 423, 725

\bibitem{4}
Chang, P., \& Bildsten, L. 2003, ApJ, 585, 464

\bibitem{5}
Chang, P., \& Bildsten, L. 2004, ApJ, 605, 830

\bibitem{6}
Chang, P., Arras, P., \& Bildsten, L. 2004, ApJ, 616, L147

\bibitem{7}
Haberl, F. 2004, Mem.\ Soc.\ Astron.\ Ital., 75, 454

\bibitem{7a}
Hailey, C.\ J., \& Mori, K. 2002, ApJ, 578, L133

\bibitem{7b}
Ho, W.\ C.\ G., \& Lai, D. 2004, ApJ, 607, 420

\bibitem{7c}
Ho, W.\ C.\ G., Lai, D., Potekhin, A.\ Y., \& Chabrier, G. 2003,
ApJ, 599, 1293

\bibitem{8}
Mereghetti, S., De Luca, A., Caraveo, P.\ A., Becker, W., Mignani, R., \& Bignami, G.\ F. 2002, ApJ, 581, 1280

\bibitem{9}
Mori, K., \& Hailey, C.\ J. 2002, ApJ, 564, 914

\bibitem{10}
Mori, K., Chonko, J.\ C., \& Hailey, C.\ J. 2004, submitted to ApJ (astro-ph/0407369)

\bibitem{11}
Pavlov, G.\ G., \& Potekhin, A.\ Y. 1995, ApJ, 450, 883

\bibitem{12}
Potekhin, A.\ Y. 1994, J.\ Phys.\ B, 27, 1073 

\bibitem{13}
Ruder, H., Wunner, G., Herold, H., \& Geyer, F. 1994, Atoms in Strong Magnetic
Fields (Springer: Berlin)

\bibitem{14}
Sanwal, D., Pavlov, G.\ G., Zavlin, V.\ E., \& Teter, M.\ A. 2002,
ApJ, 574, L61

\bibitem{8a}
Schmelcher, P., \& Cederbaum, L.\ S. 1991, Phys.\ Rev.\ A, 43, 287

\bibitem{15}
Thompson, C. \& Duncan, R.\ C. 1995, MNRAS, 275, 255 

\bibitem{15a}
Turbiner, A.\ V., \& L\'opez Vieyra, J.\ C. 2004, Modern Phys.\ Lett.\ A, 19, 1919

\bibitem{17}
van Kerkvijk, M.\ H., Kaplan, D.\ L., Durant, M., Kulkarni, S.\ R., \& Paerels, F.
2004, ApJ, 608, 432 

\bibitem{18}
Woods, P.\ M., \& Thompson, C. 2006, in Compact Stellar X-ray Sources,
ed.\ W.\ H.\ G.\ Lewin \& M.\ van der Klis (Cambridge: Cambridge
Univ.\ Press), in press (astro-ph/0406133)

\bibitem{19}
Zavlin, V.\ E., \& Pavlov, G.\ G. 2002, in Neutron Stars, Pulsars, 
and Supernova Remnants, ed.\ W.\ Becker, H.\ Lesch, \& J.\ Tr\"umper
(MPE Rep.\ 278; Garching: MPE), 263
\end{thebibliography}
\end{document}